\PassOptionsToPackage{unicode}{hyperref}
\PassOptionsToPackage{hyphens}{url}
\PassOptionsToPackage{dvipsnames,svgnames,x11names}{xcolor}
\documentclass[
  12pt]{article}
\usepackage[authoryear,round]{natbib} 
\usepackage{hyperref}

\usepackage{amsmath,amssymb}
\usepackage{iftex}
\usepackage{cleveref}
\ifPDFTeX
  \usepackage[T1]{fontenc}
  \usepackage[utf8]{inputenc}
  \usepackage{textcomp} % provide euro and other symbols
\else % if luatex or xetex
  \usepackage{unicode-math}
  \defaultfontfeatures{Scale=MatchLowercase}
  \defaultfontfeatures[\rmfamily]{Ligatures=TeX,Scale=1}
\fi
\usepackage{lmodern}
\ifPDFTeX\else  
\fi
\IfFileExists{upquote.sty}{\usepackage{upquote}}{}
\IfFileExists{microtype.sty}{% use microtype if available
  \usepackage[]{microtype}
  \UseMicrotypeSet[protrusion]{basicmath} % disable protrusion for tt fonts
}{}
\makeatletter
\@ifundefined{KOMAClassName}{% if non-KOMA class
  \IfFileExists{parskip.sty}{%
    \usepackage{parskip}
  }{% else
    \setlength{\parindent}{0pt}
    \setlength{\parskip}{6pt plus 2pt minus 1pt}}
}{% if KOMA class
  \KOMAoptions{parskip=half}}
\makeatother
\usepackage{xcolor}
\makeatletter
\ifx\paragraph\undefined\else
  \let\oldparagraph\paragraph
  \renewcommand{\paragraph}{
    \@ifstar
      \xxxParagraphStar
      \xxxParagraphNoStar
  }
  \newcommand{\xxxParagraphStar}[1]{\oldparagraph*{#1}\mbox{}}
  \newcommand{\xxxParagraphNoStar}[1]{\oldparagraph{#1}\mbox{}}
\fi
\ifx\subparagraph\undefined\else
  \let\oldsubparagraph\subparagraph
  \renewcommand{\subparagraph}{
    \@ifstar
      \xxxSubParagraphStar
      \xxxSubParagraphNoStar
  }
  \newcommand{\xxxSubParagraphStar}[1]{\oldsubparagraph*{#1}\mbox{}}
  \newcommand{\xxxSubParagraphNoStar}[1]{\oldsubparagraph{#1}\mbox{}}
\fi
\makeatother

\usepackage{longtable,booktabs,array}
\usepackage{calc} % for calculating minipage widths
\usepackage{etoolbox}
\makeatletter
\patchcmd\longtable{\par}{\if@noskipsec\mbox{}\fi\par}{}{}
\makeatother
\IfFileExists{footnotehyper.sty}{\usepackage{footnotehyper}}{\usepackage{footnote}}
\makesavenoteenv{longtable}
\usepackage{graphicx}
\makeatletter
\def\maxwidth{\ifdim\Gin@nat@width>\linewidth\linewidth\else\Gin@nat@width\fi}
\def\maxheight{\ifdim\Gin@nat@height>\textheight\textheight\else\Gin@nat@height\fi}
\makeatother
\setkeys{Gin}{width=\maxwidth,height=\maxheight,keepaspectratio}
\makeatletter
\def\fps@figure{htbp}
\makeatother

\makeatletter
\@ifpackageloaded{caption}{}{\usepackage{caption}}
\AtBeginDocument{%
\ifdefined\contentsname
  \renewcommand*\contentsname{Table of contents}
\else
  \newcommand\contentsname{Table of contents}
\fi
\ifdefined\listfigurename
  \renewcommand*\listfigurename{List of Figures}
\else
  \newcommand\listfigurename{List of Figures}
\fi
\ifdefined\listtablename
  \renewcommand*\listtablename{List of Tables}
\else
  \newcommand\listtablename{List of Tables}
\fi
\ifdefined\figurename
  \renewcommand*\figurename{Figure}
\else
  \newcommand\figurename{Figure}
\fi
\ifdefined\tablename
  \renewcommand*\tablename{Table}
\else
  \newcommand\tablename{Table}
\fi
}
\@ifpackageloaded{float}{}{\usepackage{float}}
\floatstyle{ruled}
\@ifundefined{c@chapter}{\newfloat{codelisting}{h}{lop}}{\newfloat{codelisting}{h}{lop}[chapter]}
\floatname{codelisting}{Listing}

\makeatother
\makeatletter
\@ifpackageloaded{caption}{}{\usepackage{caption}}
\@ifpackageloaded{subcaption}{}{\usepackage{subcaption}}
\makeatother

\ifLuaTeX
  \usepackage{selnolig}  % disable illegal ligatures
\fi
\usepackage[]{natbib}
\usepackage{bookmark}

\IfFileExists{xurl.sty}{\usepackage{xurl}}{} % add URL line breaks if available
\hypersetup{
  pdftitle={Title},
  pdfauthor={Author 1; Author 2},
  pdfkeywords={3 to 6 keywords, that do not appear in the title},
  colorlinks=true,
  linkcolor={blue},
  filecolor={Maroon},
  citecolor={Blue},
  urlcolor={Blue},
  pdfcreator={LaTeX via pandoc}}

\newcommand{\anon}{1}

\begin{document}

\def\spacingset#1{\renewcommand{\baselinestretch}%
{#1}\small\normalsize} \spacingset{1}

\title{\bf HR-VILAGE-3K3M: A Human Respiratory Viral Immunization Longitudinal Gene Expression Dataset for Systems Immunity}

\newif\ifanonymous
\anonymoustrue   % <-- set to \anonymousfalse for final submission

  \author{
  Xuejun Sun$^{1}$, Yiran Song$^{1}$, Xiaochen Zhou$^{1}$, Ruilie Cai$^{2}$, Yu Zhang$^{1}$,\\
  Xinyi Li$^{1}$, Rui Peng$^{1}$, Jialiu Xie$^{1}$, Yuanyuan Yan$^{1}$, Muyao Tang$^{1}$,\\
  Lakshmanane Premkumar$^{4}$, Baiming Zou$^{1}$, James S. Hagood$^{3}$,\\
  Raymond J. Pickles$^{4}$, Didong Li$^{1}$, Fei Zou$^{1,*}$, and Xiaojing Zheng$^{3,*}$\\[8pt]
  $^{1}$Department of Biostatistics\\
  University of North Carolina at Chapel Hill, Chapel Hill, NC 27514, USA\\[4pt]
  $^{2}$Department of Epidemiology and Biostatistics\\
  University of South Carolina, Columbia, SC 29208, USA\\[4pt]
  $^{3}$Department of Pediatrics\\
  University of North Carolina at Chapel Hill, Chapel Hill, NC 27514, USA\\[4pt]
  $^{4}$Department of Microbiology and Immunology\\
  University of North Carolina at Chapel Hill, Chapel Hill, NC 27514, USA\\[8pt]
  $^{*}$Corresponding authors: feizou@email.unc.edu, xiaojinz@email.unc.edu
  }

\maketitle

\if0\anon
{
  \bigskip
  \bigskip
  \bigskip
  \begin{center}
    {\LARGE\bf Title}
\end{center}
  \medskip
} \fi

\bigskip
\begin{abstract}
Respiratory viral infections pose a global health burden, yet the cellular immune mechanisms underlying protection and pathology remain unclear. Natural infection cohorts often lack pre-exposure baselines and time-controlled sampling, whereas inoculation and vaccination trials generate well-structured longitudinal transcriptomic data. However, these datasets are scattered across repositories and processed inconsistently, hindering integrative and AI-driven analyses.
To address these challenges, we developed the Human Respiratory Viral Immunization LongitudinAl Gene Expression (HR-VILAGE-3K3M) repository: an AI-ready resource integrating bulk and single-cell transcriptomic profiles from 3,178 subjects across 66 studies. The dataset spans vaccination, inoculation, and mixed exposures, with samples from blood and nasal swabs collected from public repositories including GEO, ImmPort, and ArrayExpress.
We curated and harmonized subject-level metadata, standardized outcome measures, and applied unified preprocessing with rigorous quality control. We further provide benchmark analyses illustrating its utility. This resource supports discovery of biomarkers, immune mechanisms, and methodological development. As one of the largest longitudinal transcriptomic resources for human respiratory viral immunization, HR-VILAGE-3K3M enables reproducible and scalable analyses to accelerate vaccine and antiviral research.

\end{abstract}

\noindent%
{\it Keywords:} Human respiratory viral inoculation and vaccination studies, Transcriptomic data repository, Curated subject-level metadata, Data harmonization, AI-driven analyses, Systems-immunology applications
\vfill

\newpage
\spacingset{1.8} % DON'T change the spacing!

\section{Background and Summary}

Acute respiratory viral infections, such as those caused by influenza, respiratory syncytial virus (RSV), human rhinovirus (HRV) and SARS-CoV-2, represent a significant global health burden~\citep{bender2024global,chow2023effects,hodinka2016respiratory,simoes2015challenges}. Vaccination remains the most effective strategy for preventing infection and reducing transmission~\citep{gracca2023both,yaugel2022role,nehme2011relative}. Both humoral (antibody) and cellular immunity are essential for controlling viral infections . The diversity and specificity of antibody responses are influenced by the complex interactions among various immune cells. There is a critical need to deeply understand the human cellular immune responses to antigens over time. 

However, natural infection studies frequently lack pre-infection baseline data and cannot control infection timing. Conversely, human inoculation and vaccine studies offer unique opportunities to observe immune cell responses within a structured timeline. Studies involving blood leukocytes and nasal swabs have demonstrated that both systemic and local cellular immunity are crucial in defending against respiratory viral infections and diseases~\citep{liu2016individualized,zaas2009gene,huang2011temporal,davenport2015transcriptomic,rosenheim2023sars,proud2008gene,muller2017development,habibi2020neutrophilic}. Extensive transcriptomic data from human inoculation and vaccination studies have been deposited in publicly accessible databases, such as GEO (Gene Expression Omnibus), ImmPort Shared Data, and the ArrayExpress Collection at EMBL-EBI~\citep{barrett2012ncbi,bhattacharya2018immport,parkinson2007arrayexpress}. However, the analytical power of a single cohort is limited and likely to yield inconsistent results, highlighting the need for integrated data analysis across multiple studies. Furthermore, integrating these data is invaluable for promoting reusable research and for developing advanced Artificial Intelligence (AI) and machine learning (ML) models. By leveraging the structured and detailed information from these studies, researchers can build reliable and robust transfer learning models for enhancing our understanding of infectious diseases and outcome predictions, and improving the development of personalized treatments and new vaccines. The integration of AI/ML in this domain holds the potential to revolutionize how we approach and manage infectious diseases, leading to better health outcomes ultimately.

Integrating data across studies presents several challenges. First, while public repositories host extensive gene expression datasets, there remains a significant gap between data availability and readiness for analysis. For instance, GEO provides data at various levels of processing, including raw, partially processed, and fully normalized data. These datasets are often generated through distinct study-specific pipelines by different research groups, leading to apparent batch effects. Second, metadata, including information on outcomes (e.g., infection status, symptoms, and antibody responses), is often fragmented in the supplementary files of the original papers or other places rather than where the gene expression data are deposited. For example, expression profiles are often available in GEO, while phenotypic data are stored in ImmPort, with subject-level antibody titers frequently missing. More problematic, the metadata use sample IDs that differ from those used by the expression data in several instances, making data merging impossible and requiring significant human effort to communicate with the principal investigators of those studies~\citep{grewe2024mva,lee2022prior,furman2017expression}.  Third, data have been profiled using various platforms (microarray, bulk RNA sequencing (RNA-Seq), and Single-cell RNA sequencing (scRNA-seq)) and are represented using inconsistent gene symbols due to different probe designs and methods of gene identification, such as alias gene symbols found in literature instead of official nomenclature~\citep{ziemann2016gene}. Lastly, a rigorous data preprocessing evaluation is lacking. Some datasets show abnormal sample correlations, including negative values indicative of technical artifacts, or irregular count distributions such as expression values restricted to a few discrete levels—likely stemming from improper preprocessing of raw data. In addition, duplicate samples originating from the same subjects but deposited under different studies are occasionally present, necessitating careful detection and removal to ensure consistency and integrity of the integrated dataset.

To overcome these barriers, we created the Human Respiratory Viral Immunization Longitudinal Gene Expression (HR-VILAGE-3K3M) data repository, a curated benchmark dataset encompassing over 3,000 participants from 66 studies, including vaccination, inoculation, and post-vaccination infection studies. We undertook extensive efforts to align metadata and gene expression profiles from disparate sources, ensuring a unified and analyzable structure across all datasets. This involved resolving inconsistencies in sample identifiers, correcting formatting errors (e.g., Excel-induced gene symbol corruption), and harmonizing gene annotations to official HUGO Gene Nomenclature Committee (HGNC) symbols~\citep{bruford2020guidelines}. In many cases, essential subject-level antibody data were missing from public repositories and had to be obtained directly from study authors through targeted email outreach. We also standardized heterogeneous outcome definitions across studies to enable meaningful comparisons. To ensure data integrity, we conducted rigorous quality control by identifying and excluding duplicated samples, reprocessing data from raw files when available, and filtering out datasets exhibiting abnormal or unreliable sample-level correlations. These curation and preprocessing efforts were critical in transforming raw, fragmented datasets into harmonized, ready-to-use gene expression matrices, allowing researchers to bypass the otherwise time-consuming and technically challenging steps of data discovery, cleaning, and integration.

HR-VILAGE-3K3M supports a wide range of applications, such as the discovery of early biomarkers, identification of disease-driving or causal genes, and understanding the temporal dynamics of immune responses to various pathogens through integrative data analysis across multiple studies. HR-VILAGE-3K3M will help facilitate the development of advanced AI/ML models, such as foundation models and multimodality learning, and better harness the power of AI to gain insights into infectious diseases. Ultimately, this will inform early disease diagnosis, vaccine development, and treatment strategies, promoting more robust and reproducible immunological research. In addition, it offers a benchmark database for the development of computational methodologies, such as batch effect correction, gene expression imputation, deep learning prediction, and single cell transcriptomic-informed deconvolution of bulk data.

\section{Related Work}

\subsection{Existing Datasets on Respiratory Viral Transcriptomics}
Several community-curated resources have been developed to support transcriptomic research on respiratory viruses such as influenza and COVID-19. For example, the GitHub repository \texttt{urmi-21/COVID-19-RNA-Seq-datasets} provides study-level summary matrices but does not include fully processed data or integration-ready formats for comparative analyses, and its scope is limited to COVID-19 RNA-seq studies covering infection and vaccination.

The study by Rogers et al.~\citep{rogers2019} conducted a meta-analysis of 18 publicly available influenza-related microarray datasets spanning infection, vaccination, and control groups. Their analysis identified differentially expressed genes that distinguished infection from vaccination responses and further explored variability attributable to age and sex. While this collection provides valuable biological insights, it is restricted to microarray data, lacks RNA-seq and single-cell transcriptomic studies, does not include clinical outcomes, and does not provide a unified harmonization framework across studies.

Another important resource is the pan-vaccine study by Fourati et al.~\citep{fourati2022}, which profiled immune responses across individuals vaccinated against 13 viral and bacterial pathogens. While valuable, this dataset includes only a limited number of respiratory viral vaccine samples (approximately 300 subjects), is restricted to bulk transcriptomic data, and does not provide ready-to-use processed data for downstream analyses. 

In comparison, \textbf{HR-VILAGE-3K3M} provides the most comprehensive and harmonized resource to date for respiratory viral studies. It integrates microarray, bulk RNA-seq, and single-cell RNA-seq datasets across 66 studies, covering over 3,000 subjects and 14,000 profiles. Importantly, vaccine and challenge studies in HR-VILAGE-3K3M provide defined pre- and post-exposure measurements along controlled timelines, enabling direct investigation of immune dynamics.

\subsection{Batch Effect Correction and Data Harmonization Methods}
A critical challenge in building integrated multi-study resources is mitigating batch effects. For microarray data, commonly used approaches include ComBat (empirical Bayes adjustment)~\citep{wang2017combat}, limma (linear modeling with moderated variance)~\citep{ritchie2015limma}, and the sva framework (surrogate variable analysis)~\citep{leek2012sva}. For RNA-seq data, extensions such as ComBat-seq~\citep{zhang2020combatseq} and svaseq~\citep{leek2014svaseq} have been developed.  

Single-cell RNA-seq integration presents unique challenges due to sparsity and high dimensionality. Widely used tools include Harmony (iterative alignment in PCA space)~\citep{korsunsky2019harmony}, canonical correlation analysis (CCA)~\citep{stuart2019comprehensive}, and mutual nearest neighbor (MNN) mapping~\citep{haghverdi2018mnn}.  

Cross-platform integration methods bridging microarray and RNA-seq include Quantile Normalization (QN)~\citep{bolstad2003}, Training Distribution Matching (TDM)~\citep{thompson2015}, and Bayesian approaches leveraging control samples to estimate batch effects~\citep{tang2021rankin}. While QN was originally designed for microarray normalization, it is often applied to RNA-seq in practice due to its robustness across platforms.  

Ultimately, the choice of integration method depends on the modality, study design, and downstream objectives. In this work, we benchmark multiple normalization strategies and adopt harmonization approaches that maximize reproducibility and biological interpretability across diverse datasets.

\section{Methods}

\begin{figure}[!h]
  \centering
  \includegraphics[width=\textwidth]{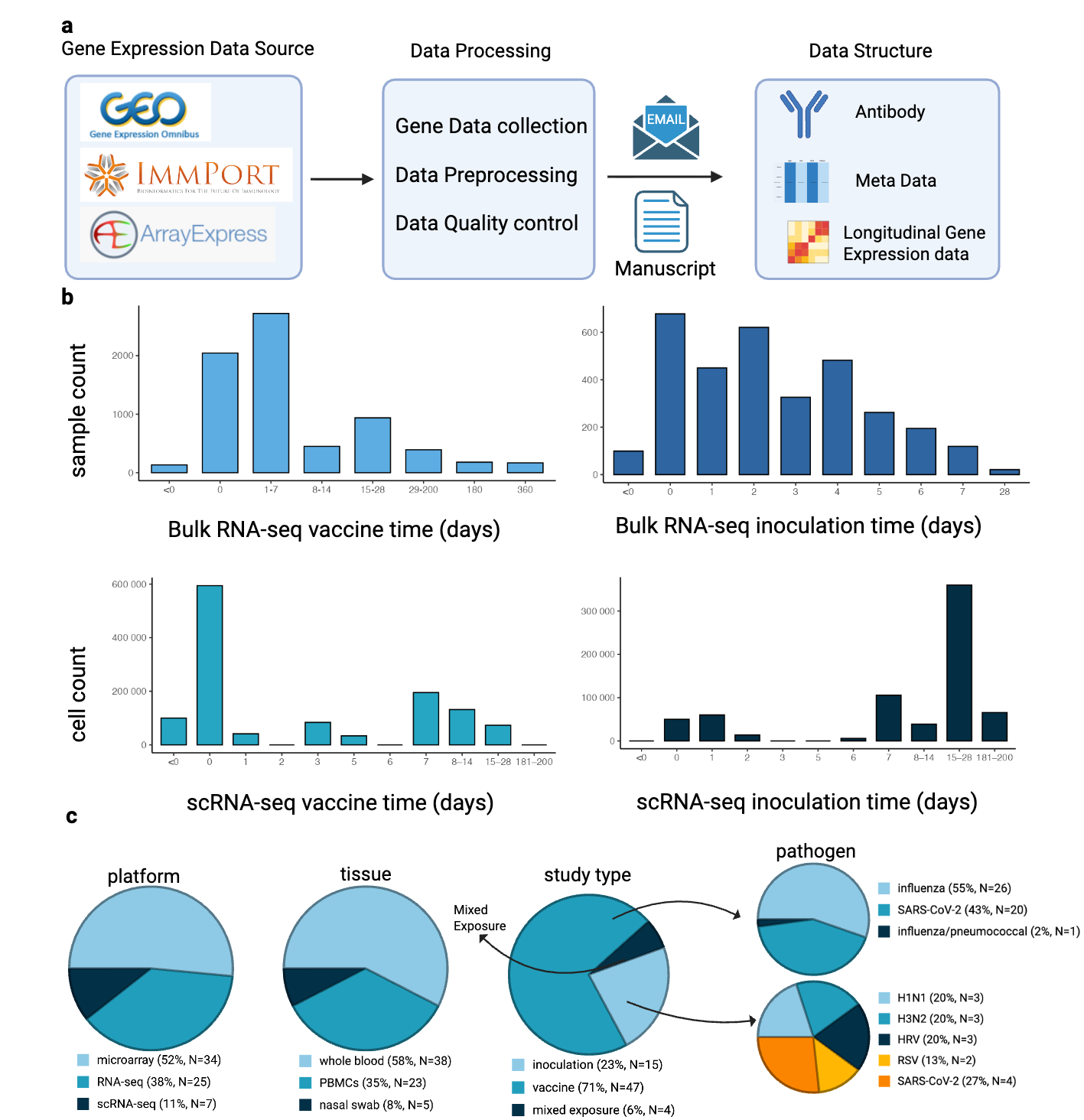}
  \caption{Overview of HR-VILAGE-3K3M. (a) HR-VILAGE-3K3M construction workflow. (b) Distribution of sample timepoints for vaccine and inoculation studies, shown separately for bulk RNA-seq and single-cell RNA-seq datasets. (c) Composition of the dataset, stratified by platform, tissue type, study type, and pathogen, including both bulk and single-cell transcriptomic studies.}
  \label{fig:rigel}
\end{figure}

%\subsection{Data collection and curation}
\noindent\textbf{Data collection and curation.}
The data collection workflow of HR-VILAGE-3K3M is summarized in Figure \ref{fig:rigel}a. HR-VILAGE-3K3M consists of three components: gene expression profiles, metadata, and raw antibody data. We previously established a collection of human blood and nasal swab gene expression datasets from viral respiratory inoculation and vaccination studies. These datasets were gathered using targeted search queries from three publicly available repositories: GEO, ImmPort, and ArrayExpress, as detailed in~\citep{lan2024curated}. HR-VILAGE-3K3M extends the gene expression profiles from our previous collection by incorporating newly acquired data from recent studies. The metadata was obtained from GEO, ImmPort, and supplementary materials. Subject-level antibody titers were obtained from GEO, ImmPort, supplementary files, and by requesting them from study authors via email. 

%\subsection{Data processing}
\noindent\textbf{Data processing.}
For bulk expression data, the profiling platforms and data processing states vary greatly across GEO (microarray and RNA-seq), ImmPort (microarray), and ArrayExpress (RNA-seq). GEO requires raw FASTQ files for RNA-seq but does not mandate raw microarray data. Advantageously, all Affymetrix microarray studies in HR-VILAGE-3K3M have raw CEL files. We thus processed them using a standardized pipeline that includes \texttt{RMA}  normalization \citep{irizarry2003summaries}, probe-to-gene symbol mapping, Quantile Normalization (QN, \citep{bolstad2003comparison}), and $\log_2(x)$ transformation. When multiple probes mapped to the same gene, the probe with the highest median expression was retained. For Illumina microarray data, raw IDAT files were rarely available, but most studies provided non-normalized intensity data, which were processed by mapping probes to gene symbols using Illumina annotation tables, followed by QN and $\log_2(x)$ transformation. For RNA-seq, we mapped Ensembl IDs to HGNC gene symbols and provided raw counts data. In cases where only FASTQ files were available and raw counts were not provided, we performed read alignment and quantification to generate count matrices. All final expression matrices are provided at the gene level, with raw probe-level matrix data also available (when applicable) to support custom preprocessing workflows. 

For scRNA-seq, the studies provided variable data formats, including h5ad files, 10x matrix, and RObject files, with or without the metadata. We thus first integrate the metadata into the Python AnnData object \citep{Virshup2024anndata}, merge the data from the same study, and align to have consistent gene identifiers and cell-level metadata. Then we process them with the standardized preprocessing pipeline, including removal of genes expressed in fewer than 10 cells, total-count normalization (target sum: 10,000), and \texttt{log1p} transformation. A dataset label was added to observations for tracking provenance. All scRNA-seq datasets were saved in AnnData (.h5ad) format, as it has an efficient processing speed and relatively smaller size. For visualization, all datasets were concatenated with an inner join over a shared set of 13,589 genes, with normalization, \texttt{log1p} transformation, Principal Component Analysis (PCA), neighborhood graph construction, and Uniform Manifold Approximation and Projection (UMAP,~\citealp{mcinnes2018umap}) embedding.

%\subsection{Data quality control}
\noindent\textbf{Data quality control.}
Across all platforms, we observed variability in how gene identifiers were handled—some studies had already mapped probes or Ensembl IDs to gene symbols, while others had not. Even among those that did, gene symbol formatting was inconsistent, some with outdated or even misformatted entries (e.g., “14-Sep” instead of SEPTIN14) due to autoformatting by Excel~\citep{ziemann2016gene}. To ensure consistency and accuracy, we standardized all gene symbols using the \texttt{HGNChelper} R package~\citep{oh2022hgnchelper}, aligning them with current HGNC-approved nomenclature and removing entries that did not match official symbols.

We also identified duplicate samples in different GEO datasets, which occurred when datasets are reused across multiple publications or submitted to GEO more than once~\citep{liu2016individualized,woods2013host,huang2011temporal,zaas2009gene}. We removed duplicate samples derived from the same subjects. Potential duplicates were flagged by reviewing study protocols and confirmed using Spearman correlation of gene expression profiles. For datasets originating from the same study group, we retained only the version with the most comprehensive timepoint coverage per subject and documented the duplicate study information on our data website to ensure transparency and traceability.

We used GEO accession numbers as unique identifiers for each observation, ensuring that the row names of the metadata aligned with the row names of the gene expression matrices. A consistent subject identifier was also defined, allowing researchers to accurately link antibody measurements with metadata across studies. For gene-level quality control, probe IDs were mapped to gene symbols using platform-specific annotation tables from GEO, while Ensembl IDs were converted using the \texttt{biomaRt} package in R~\citep{bruford2020guidelines}. The resulting gene symbols were then standardized to official HGNC symbols using the \texttt{HGNChelper} R package~\citep{oh2022hgnchelper}. 

Additionally, for scRNA-seq, the primary expression matrix was verified or reassigned from raw count layers (\texttt{raw\_counts}, \texttt{counts}, or equivalent). Sample-specific cell barcodes were prefixed with dataset identifiers to ensure uniqueness. All metadata fields of interest—such as donor ID, sample ID, time point (day), COVID status, sex, age, tissue type, and study type—were standardized across datasets. Missing columns were added where necessary and initialized with NA values, ensuring a consistent structure suitable for downstream integration and analysis. Duplicate metadata columns were removed to prevent merge conflicts. These steps ensured a harmonized, high-quality dataset suitable for downstream integration and analysis.

%\subsection{Outcome definition}
\noindent\textbf{Outcome definition.}
Different studies often have varying objectives and outcome definitions, making cross-study comparisons challenging. To address this, we established consistent outcome criteria across studies. For inoculation studies, infection status (infected vs.\ uninfected) and symptom status (symptomatic vs.\ asymptomatic) were typically predefined in the original publications~\citep{liu2016individualized,habibi2020neutrophilic,proud2008gene,davenport2015transcriptomic,muller2017development}. For influenza vaccine studies, we defined response status based on the Maximum Fold Change (MFC) in Hemagglutination Inhibition (HAI) titers: high responders were defined as having MFC~$\geq$~4 and a Day~28 HAI titer~$\geq$~40, while non-responders had MFC~$\leq$~1~\citep{hipc2017multicohort}. Participants who did not meet either threshold were classified as moderate responders. Raw antibody measurements for influenza studies are included in the \texttt{antibody} folder for reference and further analysis. For COVID-19 vaccine studies, where no universal response definition exists, we provided binding and neutralizing antibody data in the antibody folder, allowing users to define customized outcome criteria tailored to their research needs.

%\subsection{Summary}
\noindent\textbf{Summary.}
We generated a rigorously processed meta-source benchmark expression dataset, HR-VILAGE-3K3M, comprising 3{,}178 participants and 14{,}136 expression profiles from 66 studies. It includes 462 participants (18–55 years) from 15 inoculation studies involving five viruses (H3N2, H1N1, HRV, RSV, and SARS-CoV-2), profiled at 1–23 time points. Additionally, it includes 2{,}412 participants (0.5–89+ years) from 47 influenza and COVID-19 vaccination studies, profiled at 1–22 time points, and 304 participants from four studies with mixed exposure, profiled at 2–7 time points. Figure \ref{fig:rigel}b shows the bulk RNA-seq and scRNA-seq distribution of sample collection times following vaccination and inoculation, with dense sampling in the first few days for inoculation, while vaccination samples span longer durations—up to 180 and 360 days—though at lower frequencies. For each study, we provided both processed and raw gene expression data, along with the corresponding metadata and subject-level antibody information. Regarding sample types, 38 studies (58\%) used whole blood, 23 (35\%) used human peripheral blood mononuclear cells (PBMCs), and five (8\%) used nasal swabs. Of these studies, 34 (52\%) were conducted using microarray platforms, 25 (38\%) using RNA-seq, and seven (11\%) using 10x Genomics. In terms of study type, 15 (23\%) were inoculation studies, 47 (71\%) were vaccine studies, and 4 (6\%) were mixed exposure studies, including post-vaccination infection, post-infection vaccination, and natural infection (Figure \ref{fig:rigel}c). HR-VILAGE-3K3M provides a user-friendly platform for researchers to analyze the temporal dynamics of systems immunity to viral infections. The data is hosted on Hugging Face, enabling efficient access through Python with support for parallelized loading and selective retrieval. We also provide custom data splits for flexible loading by study type, tissue, and more. Our dataset is publicly available on Hugging Face at \url{https://huggingface.co/datasets/xuejun72/HR-VILAGE-3K3M}.

\section{Technical Validation}
In this section, we illustrate the utility of HR-VILAGE-3K3M through three representative analyses: outcome prediction using longitudinal bulk data and cell type annotation using paired bulk and single-cell data. 

\subsection{Age and Sex Prediction Using Longitudinal Bulk Data}

We evaluated how well baseline whole-blood transcriptomes capture two fundamental biological traits—sex and age—using microarray and RNA-seq datasets from HR-VILAGE-3K3M. Sex prediction used a literature-curated set of Y-chromosome marker genes \citep{ellis2018sex}, while age prediction leveraged genome-wide expression features within a cross-validation framework. These analyses served two purposes: (i) providing a robust quality-control (QC) screen for detecting potential sample mislabeling, and (ii) establishing a quantitative benchmark for predictive modeling across transcriptomic platforms.

The analyses for sex prediction included 1{,}096 microarray and 646 RNA-seq baseline samples. Expression matrices were first quantile-normalized (QN) across all genes \citep{bolstad2003comparison}, and RNA-seq data were further transformed using $\log_2(x+1)$ to stabilize variance \citep{law2014voom}. Y-chromosome marker genes were then selected from the normalized data, retaining only those expressed in all samples to ensure consistent cross-platform coverage. For each sample, a composite Y-score was computed as the sum of normalized expression values across all selected Y-chromosome marker genes. The optimal threshold separating males and females was determined using Youden’s $J$ statistic\citep{youden1950index}, maximizing the difference between true-positive and false-positive rates (TPR – FPR) on the ROC curve. Samples with Y-scores above this threshold were classified as male, while those below were classified as female.

The analyses for age prediction included 1{,}116 microarray and 784 RNA-seq baseline samples. Expression matrices were quantile-normalized (QN) \citep{bolstad2003comparison}, and RNA-seq data were transformed using $\log_2(x+1)$ to stabilize variance \citep{law2014voom}. Genes missing in any sample were removed, and low-information genes (bottom 25\% by mean and variance) were excluded to improve model stability. Age (in years) was used as the target variable. Within each cross-validation fold, the top 2{,}000 highly variable genes (HVGs) were selected from the training data \citep{yip2017evaluation}, two modeling strategies were applied: (1) Ridge regression with L2 regularization optimized via \texttt{RidgeCV} \citep{hoerl1970ridge,scikit-learn}, and (2) a Fully Connected Neural Network (FCNN) comprising two hidden layers (512 and 256 ReLU units), a dropout rate of 0.3, and the Adam optimizer (learning rate $1\times10^{-3}$) \citep{chollet2015keras,kingma2014adam}. Early stopping and learning rate scheduling were employed to mitigate overfitting. Models were evaluated using 10-fold cross-validation with out-of-fold predictions \citep{arlot2010survey}. Performance metrics included the coefficient of determination ($R^2$) and mean absolute error (MAE, years), providing a consistent framework for assessing both linear and nonlinear age-related expression patterns across modalities.

\begin{table}[ht]
\centering
\caption{Summary of Y-chromosome marker panels and classification performance.}
\begin{tabular}{lccc}
\hline
\textbf{Dataset} & \textbf{Number of Markers} & \textbf{Threshold} & \textbf{ROC-AUC} \\
\hline
Microarray & 11 & 76.2183 & 0.989 \\
RNA-seq & 9 & 33.4438 & 0.991 \\
\hline
\end{tabular}
\label{tab:sex_marker_auc}
\end{table}

\begin{table}[ht]
\centering
\caption{Age prediction performance across data modalities and models using top-2000 highly variable genes (HVGs). Values show mean (95\% CI) across 5 CV folds.}
\begin{tabular}{lccc}
\hline
\textbf{Dataset} & \textbf{Model} & \textbf{$R^2$ (95\% CI)} & \textbf{MAE (years, 95\% CI)} \\
\hline
Microarray & FCNN & 0.855 (0.827, 0.882) & 6.47 (6.15, 6.79) \\
RNA-seq & FCNN & 0.704 (0.636, 0.773) & 6.88 (6.30, 7.47) \\
Microarray & Ridge Regression & 0.830 (0.806, 0.854) & 7.84 (7.59, 8.10) \\
RNA-seq & Ridge Regression & 0.674 (0.575, 0.773) & 7.56 (6.90, 8.22) \\
\hline
\end{tabular}
\label{tab:age_prediction}
\end{table}

Across both data modalities, baseline whole-blood transcriptomes captured strong and biologically consistent signals for both sex and age. The Y-chromosome marker--based classifiers achieved near-perfect discrimination between males and females, with ROC-AUC values of 0.989 for the microarray and 0.991 for the RNA-seq datasets (Table 1). For age prediction, Ridge regression models achieved \(R^2 = 0.829\) (MAE = 7.90 years) for microarray and \(R^2 = 0.708\) (MAE = 7.26 years) for RNA-seq, while the FCNN further improved performance to \(R^2 = 0.859\) and \(R^2 = 0.728\), respectively (Table 2). These results demonstrate that baseline transcriptomic profiles encode substantial age-related variation, with deep neural architectures capturing modest nonlinear gains. Microarray data yielded slightly higher predictive stability than RNA-seq, likely reflecting differences in cohort composition and platform-specific variability rather than intrinsic disparities in biological signal.

\subsection{Outcome Prediction Using Longitudinal Bulk Data}

To demonstrate the utility of the HR-VILAGE-3K3M dataset for evaluating batch effect correction and predictive modeling, we conducted a comprehensive benchmark analysis. We selected 19 influenza vaccine studies comprising 1,210 participants with balanced numbers of high and low antibody responders. Gene expression profiles for these participants spanned 1 to 19 time points, and we retained only samples collected during the early immune response window (prior to day 28 before antibody measurement). To stabilize variance, RNA-seq data were log-transformed using $\log_2(x + 1)$, and genes in the bottom 25\% of both mean expression and variance were removed to reduce technical noise.

We applied three batch effect correction methods---QN~\citep{bolstad2003comparison}, ComBat~\citep{johnson2007adjusting}, and Regression from \texttt{scanpy~\citep{wolf2018scanpy}}---using study identifiers as the batch variable. To assess the effectiveness of these corrections, we visualized batch and outcome separation using t-SNE projections of the normalized gene expression data, highlighting clustering by both study and responder status.

For predictive modeling, we first applied a PCA + generalized estimating equations (GEE) framework with an exchangeable working correlation structure to account for within-subject dependence. Gene expression data were projected onto the top 200 principal components (PCs) from the training folds and combined with a centered and scaled time covariate as well as categorical vaccine and tissue predictors. These features were used in a binomial GEE with robust variance estimators, and subject-level outcomes were summarized by majority vote across sample-level predictions. The second strategy used deep learning models, including Recurrent Neural Network (RNN)~\citep{elman1990finding,rumelhart1986learning}, Long Short-Term Memory (LSTM)~\citep{hochreiter1997long}, Gated Recurrent Unit (GRU)~\citep{cho2014learning}, and a Transformer~\citep{vaswani2017attention}. Time-ordered gene expression sequences were constructed for each subject, padded to a fixed length, and concatenated with numeric timepoints. In the Transformer, temporal information was encoded using Fourier-based positional encodings and two Pre-Norm encoder layers with multi-head attention, feed-forward blocks, and residual connections. Encoded representations were aggregated via masked mean pooling and passed through fully connected layers for binary classification. All models were trained using binary cross-entropy loss with the AdamW optimizer~\citep{kingma2014adam} and evaluated via stratified 5-fold cross-validation repeated across six random seeds.

\begin{figure}[h]
  \centering
  \includegraphics[width=1\textwidth]{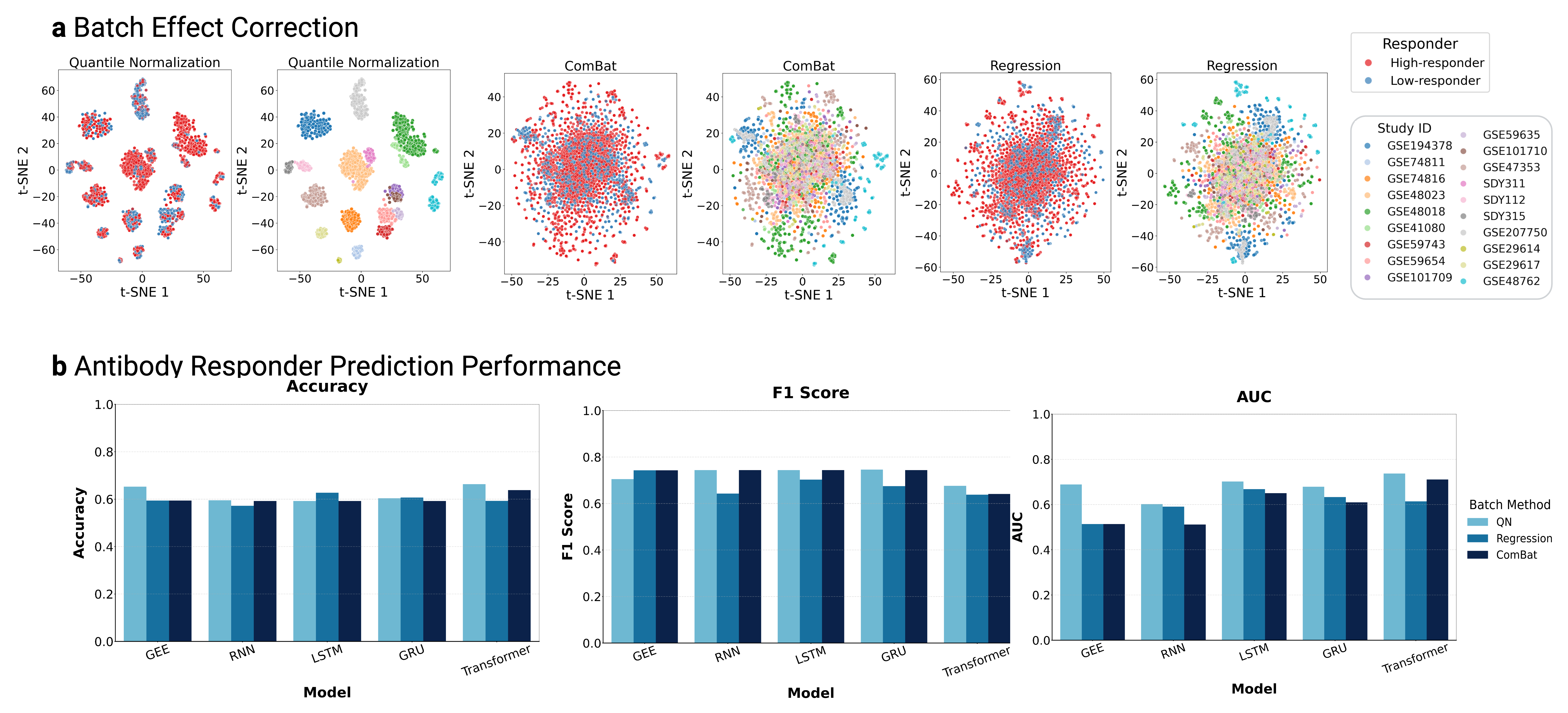}
  \caption{Evaluation of Batch Effect Correction and Predictive Modeling on the HV-RIGEL-3K Dataset. (a) Batch Effect Correction Visualization. t-SNE plots display sample clustering before and after batch effect correction using QN, Regression, and ComBat. Points are colored by responder status (left panels) or study ID (right panels) to assess preservation of biological signal and reduction of batch-specific variation. (b) Antibody Responder Prediction Performance. Bar plots show mean accuracy, AUC, and F1 score across five modeling approaches—PCA-GEE, RNN, LSTM, GRU, and Transformer—under three batch correction methods (QN, Regression, ComBat). Results are averaged across stratified 5-fold cross-validation and six random seeds. }
  \label{fig:prediction}
\end{figure}

\textbf{Batch effect correction.} The QN-corrected data show clear separation by study (indicating preserved batch effects) but also better within-study separation by outcome. Regression moderately mixes study batches while retaining some biological structure. ComBat achieves the strongest batch mixing, as indicated by overlapping study clusters, but this comes at the cost of reduced separation between responder groups (Figure \ref{fig:prediction}a).

\textbf{Antibody Responder Prediction.} 
With quantile normalization (QN), the Transformer achieved the best overall performance, reaching the highest AUC (0.737) and accuracy (0.663), underscoring its ability to capture temporal dependencies in longitudinal gene expression data. Recurrent models (RNN, LSTM, GRU) showed moderate performance, with AUCs ranging from 0.602 to 0.702, while ComBat and Regression corrections generally produced lower scores. Overall, QN correction consistently enhanced predictive performance. These findings highlight the HR-VILAGE-3K3M dataset’s value as a benchmark for evaluating models of immune response dynamics (Figure~\ref{fig:prediction}b).

\subsection{Cell Type Annotation Using Paired Bulk and Single-Cell Data}

In this analysis, we investigated immune cell-type dynamics using paired bulk RNA-seq and scRNA-seq data from SARS-CoV-2 vaccine studies, with PBMCs samples collected from the same subjects on Day 1 and Day 7 from datasets GSE201533 and GSE201534.

\begin{figure}[!h]
  \centering
  \includegraphics[width=\textwidth]{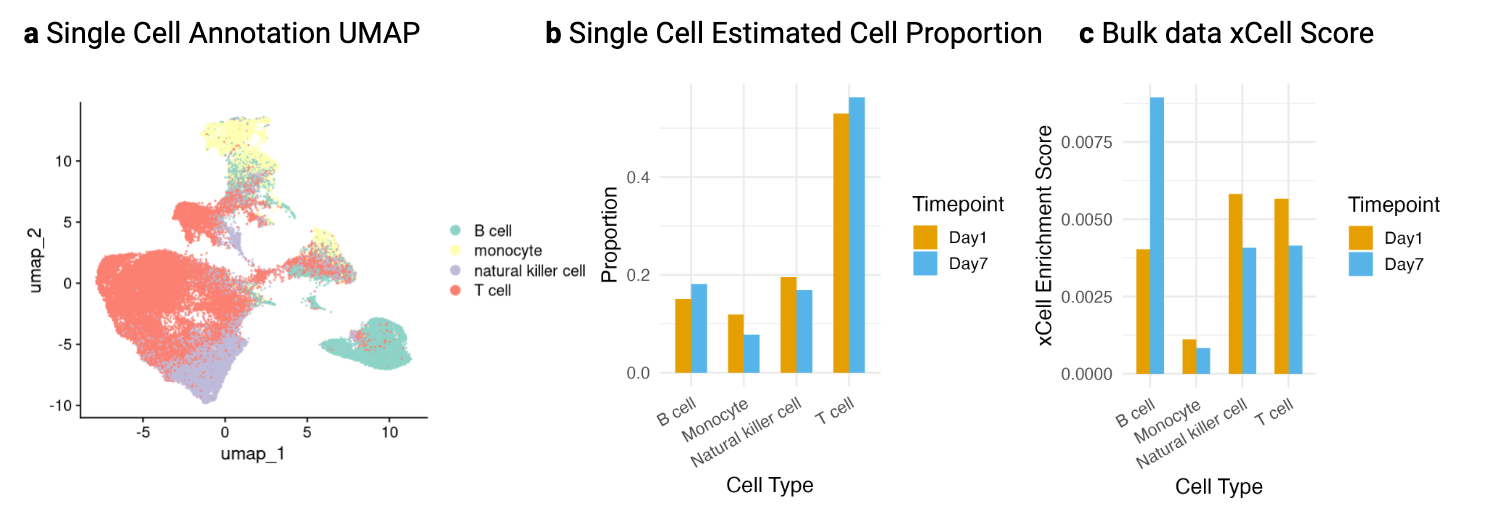}
  \caption{Comparison of immune cell-type composition and transcriptional activity using paired single-cell and bulk RNA-seq. (a) UMAP visualization of scRNA-seq data showing major immune cell types, colored by cell type annotation. (b) Bar plot of estimated immune cell-type proportions at Day 1 and Day 7 derived from single-cell RNA-seq data. (c) Bar plot of xCell enrichment scores for the same cell types based on bulk RNA-seq data at Day 1 and Day 7.}
  \label{fig:sc}
\end{figure}

For the single-cell data, since the original study lacked curated cell type annotations, we utilized two automated annotation methods, \texttt{SingleR} (a reference-based approach, \citep{aran2019reference} and \texttt{GPTCelltype} (a language-model-based approach, \citep{hou2024gptcelltype}. Upon comparison, we observed that \texttt{SingleR} frequently produced rare or biologically implausible annotations, often assigning cell types with fewer than 100 cells or labels not typically present in PBMCs. Therefore, we supplemented and refined the annotation using GPTCelltype results to ensure biological consistency. The final annotation revealed that T cells comprised the dominant immune population (over 50\%) at both time points, followed by Natural Killer (NK) cells and B cells (15–20\% each). Monocytes were detected at lower proportions (Figure \ref{fig:sc}a).

For the bulk data, we applied the xCell~\citep{aran2017xcell} algorithm to estimate enrichment scores for major immune cell types, including B cells, T cells, monocytes, and natural killer cells. Enrichment scores were computed by aggregating xCell subtype scores into cell type level, followed by averaging across samples for each time point. Unlike scRNA-seq-based proportions, xCell scores from bulk RNA-seq reflect both immune cell-type abundance and transcriptional activity, as the algorithm infers enrichment scores based on predefined gene signatures that capture cell-type-specific expression patterns. Importantly, xCell scores are not compositional and are not constrained to sum to one as they are influenced not only by cell frequency but also by the activation state of the cell populations. 

Comparison between the two modalities reveals consistent temporal trends (Figure~\ref{fig:sc}b--c). Both scRNA-seq–derived cell proportions and xCell enrichment scores indicate an increase in B cells from Day~1 to Day~7, alongside a decrease in monocytes and NK cells. In contrast, T cell levels remain relatively stable across timepoints in the scRNA-seq data, while a modest decline in xCell scores may reflect reduced transcriptional activity or shifts in T cell activation states rather than changes in abundance.

\section{Potential HR-VILAGE-3K3M Applications}

\begin{figure}[!h]
  \centering
  \includegraphics[width=\textwidth]{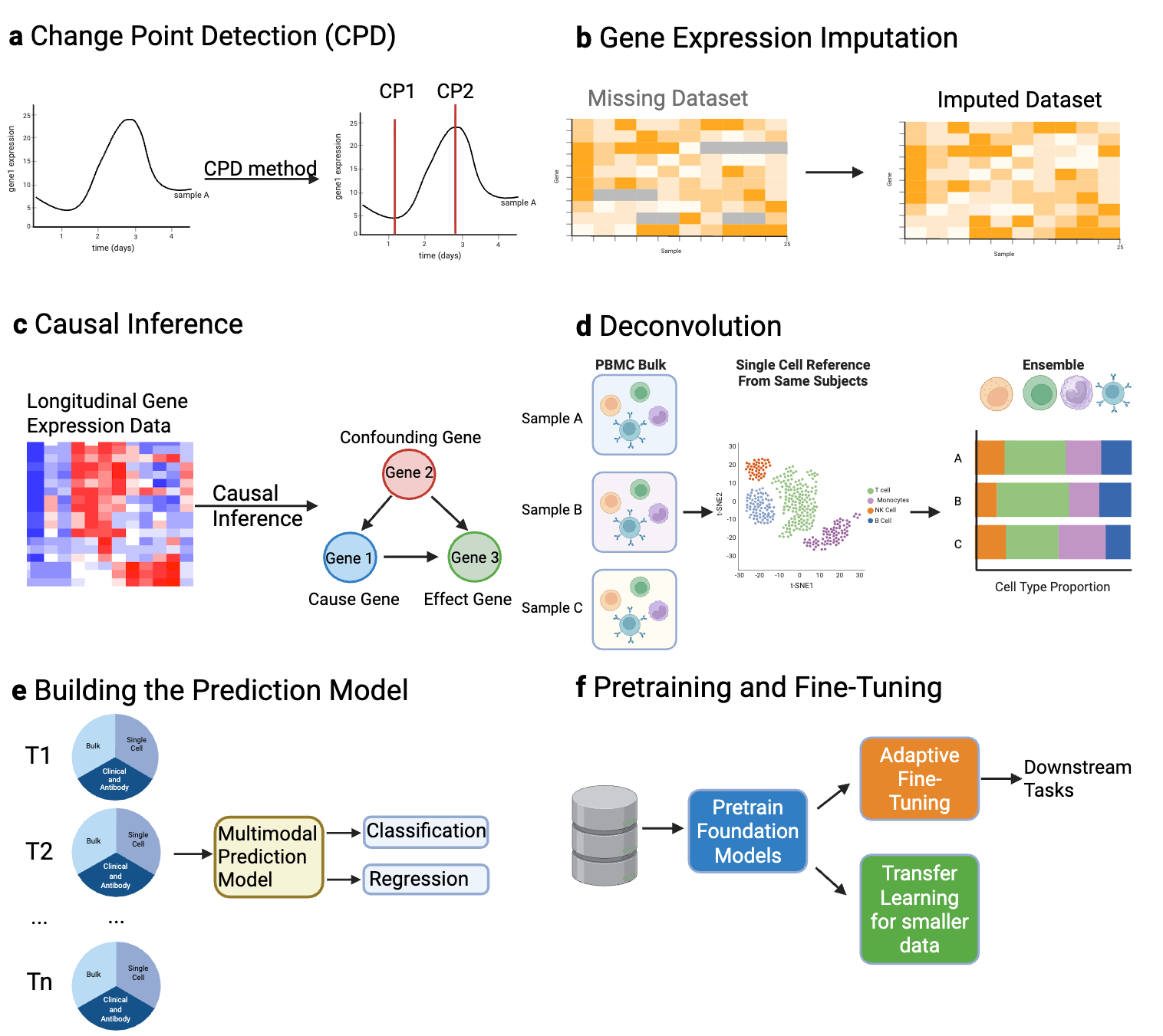}
  \caption{Potential tasks using HR-VILAGE-3K3M. (a) Change point detection. (b) Gene expression imputation. (c) Causal inference (d) Deconvolution. (e) Building prediction model. (f) Pretraining and fine-tuning. }
  \label{fig:task}
\end{figure}

\textbf{Change point detection.}
The high-resolution transcriptomic data also enable the detection of the onset of cellular or pathway activation using change point analysis. Genes that exhibit the earliest expression changes in a pathway post-immunization are more likely to be driver genes. Additionally, the characteristics of HR-VILAGE-3K3M, including irregular sampling intervals, variable timepoints, and heterogeneous individual responses, make it particularly well-suited for benchmarking change point detection algorithms (Figure \ref{fig:task}a). Furthermore, the availability of single-cell data facilitates the investigation of temporal changes in specific immune cell subsets, such as monocytes or T cells, including shifts in their abundance and pathway-level activation dynamics following immunization.

\textbf{Gene expression imputation.}
Due to platform-specific differences, the bulk transcriptomic datasets in HR-VILAGE-3K3M exhibit a distinct pattern of missingness, where gene availability is largely determined by the study and profiling platform. This deviates from the assumptions of Missing Completely at Random (MCAR) or Missing at Random (MAR), posing challenges for data integration and downstream analysis. HR-VILAGE-3K3M thus serves as a valuable benchmark for developing and evaluating imputation methods specifically designed to address this structured missingness, thereby enhancing the utility of heterogeneous transcriptomic data (Figure \ref{fig:task}b).

\textbf{Causal inference.}
Longitudinal studies are particularly effective at establishing cause-and-effect relationships. The HR-VILAGE-3K3M dataset, which contains multiple high-resolution transcriptomic data, offers valuable opportunities to develop advanced statistical methods, such as cross-lagged panel models, structural equation modeling, and dynamic causal modeling, to infer causal relationships. It enhances the validity of causal inferences by controlling for time-varying confounders and reducing biases associated with cross-sectional data. Additionally, the immunization longitudinal dataset captures temporal dynamic changes induced by immunization in the transcriptomic network structure, providing deeper insights into how immunization impacts the regulatory network (Figure \ref{fig:task}c).

\textbf{Deconvolution.}
Two studies in HR-VILAGE-3K3M contain longitudinal paired bulk and single-cell gene expression data from the same individuals. Using time-series paired samples improves the accuracy of deconvolution methods due to the consistent biological context. Additionally, paired datasets allow for cross-validation, where single-cell data can be used to validate and refine the deconvolution results obtained from bulk data (Figure \ref{fig:task}d).

\textbf{Building Prediction Model.}
Longitudinal data that integrates bulk and single-cell transcriptomics with clinical and antibody measurements provides a rich foundation for predictive modeling using multimodal deep learning. Models such as RNNs, LSTMs, GRUs, and Transformers are well-suited to capture temporal patterns and cross-modal relationships within this heterogeneous data. These approaches can support a wide range of prediction tasks, including classification (e.g., vaccine responder vs. non-responder), regression (e.g., prediction of antibody titers). By incorporating multiple biological and clinical modalities, these models enable the discovery of early immune signatures, enhance interpretability, and support applications such as risk stratification, response prediction, and personalized immunization strategies (Figure \ref{fig:task}e).

\textbf{Pretraining and Fine-Tuning.}
The scale of heterogeneity of HR-VILAGE-3K3M makes it a resource for pretraining foundation models to capture complex human immune response dynamics to vaccination and inoculation. By integrating data from a diverse set of studies across platforms, tissues, timepoints, and pathogens, the dataset provides rich biological variation that can drive generalizable feature learning. After pretraining, the model can then be adapted through adaptive fine-tuning for specific downstream tasks (outcome prediction), and transfer learning to smaller, underpowered datasets that benefit from shared temporal or biological structure (Figure \ref{fig:task}f).

Overall, HR-VILAGE-3K3M provides a large, high-quality dataset that enables a wide range of downstream analyses and methodological development in systems immunology. By integrating longitudinal gene expression datasets from diverse studies and platforms, HR-VILAGE-3K3M supports the study of dynamic immune responses over time, facilitates cross-study comparisons, and allows robust benchmarking of computational tools.

\section{Author Contribution}
Conceptualization: X.S., F.Z., X.Z.; Methodology: X.S., F.Z., X.Z.; Data curation: X.S., Y.S., X.Z., X.L., R.P., X.Z.; Data infrastructure and resource development: R.C., Data preprocessing and quality control: X.S., Y.S., Y.Z., J.X., M.T.; Data integration: X.S; Data analysis: X.S., Y.S.; Investigation: L.P., B.Z., J.S.H., R.J.P., D.L.; Validation: X.S., F.Z., X.Z.; Supervision: F.Z., X.Z.; Writing—original draft preparation: X.S., Y.S., F.Z., X.Z.; Writing—review and editing:  L.P., B.Z., J.S.H., R.J.P., D.L., F.Z., X.Z.

\section{Acknowledgment}

We sincerely thank the researchers who generously provided additional data for this work, including Simone Lucchesi, Ali H. Ellebedy, Hye Kyung Lee, Leonie M. Weskamm, Reinhold Förster, Swantje I. Hammerschmidt, Ellie N. Ivanova, Helder Nakaya, Xiaojun Li, and Ryan Thwaites. Their support significantly improved the quality and completeness of the HR-VILAGE-3K3M resource.

We also gratefully acknowledge the researchers who responded to our outreach and provided helpful clarifications, including Hildegund Ertl, John Wherry, Kai Huang, Patricia W. Finn, Octavio Ramilo, Damien Chaussabel, Ellie Ivanova, Steven Kleinstein, Hailong Meng, Darawan Rinchai, Lothar Hennighausen, Maike Buchner, Leonie M. Weskamm, Qiuyu Gong, Claire E. Gustafson, and Purvesh Khatri. We sincerely appreciate their time, insights, and willingness to engage with our team.

\section{Funding}

This work did not receive funding.

\section{Disclosure statement}\label{disclosure-statement}

The authors declare that they have no conflict of interest

\section{Data Availability Statement}\label{data-availability-statement}

The HR-VILAGE-3K3M dataset is available on Hugging Face (\href{https://huggingface.co/datasets/xuejun72/HR-VILAGE-3K3M}{\texttt{xuejun72/HR-VILAGE-3K3M}}), with analysis code accessible on both Hugging Face and GitHub (\href{https://github.com/XuejunSun98/HR-VILAGE-3K3M}{\texttt{XuejunSun98/HR-VILAGE-3K3M}}). Version~1 was released on May~17,~2025, followed by version~2 on October~13,~2025.

\clearpage
\phantomsection
\label{supplementary-material}

\section*{Supplementary Information}

\renewcommand{\thesubsection}{Supplementary~\arabic{subsection}}

\subsection{List of Studies Included in the HR-VILAGE-3K3M Dataset}
We include the full list of reference of our dataset~\citealp{lindeboom2024human,proud2008gene,davenport2015transcriptomic,muller2017development,habibi2020neutrophilic,liu2016individualized,zhai2015host,sparks2023influenza,jochems2018inflammation,nakaya2015systems,franco2013integrative,bucasas2011early,furman2013apoptosis,furman2014systems,furman2017expression,thakar2015aging,avey2020seasonal,tsang2014global,rinchai2022high,forst2022vaccination,odak2024systems,lee2022comprehensive,lee2022heterologous,lee2022mrna,lee2022prior,ryan2023systems,watanabe2023time,drury2024multi,grewe2024mva,chang2024altered,chang2024network,arunachalam2021systems,wimmers2021single,aydillo2022transcriptome,thwaites2023early,nakaya2011systems,cao2014differences,obermoser2013systems,narang2018influenza,monaco2019rna,carre2021endoplasmic,henn2013high,sarin2024impaired,huang2024interferon,ivanova2023mrna,kim2022germinal,lindeboom2023human,stephenson2024temporal,blazkova2017multicenter,alpert2019clinically,tomic2019fluprint}

\subsection{Experimental Compute Details}
Bulk data preprocessing and multimodal data alignment were performed locally using RStudio on a personal computer (Apple M4 chip, 16GB RAM, macOS). Single-cell preprocessing and antibody prediction model development were performed on an HPC cluster environment. Interactive Jupyter notebook sessions were configured with 4 CPU cores and 150 GB of RAM, without GPU acceleration. For memory-intensive operations such as merging single-cell datasets, batch jobs were submitted with increased computational resources, including up to 700 GB of RAM. Single-cell annotation was conducted in RStudio on the HPC cluster, utilizing up to 100 GB of RAM to accommodate tasks such as dimensionality reduction, clustering, and label assignment. Cell-type enrichment analysis was performed locally using the \texttt{xCell} R package, as it required minimal computational overhead. The combination of local and HPC-based computation ensured efficient execution of both lightweight preprocessing steps and computationally intensive deep learning models, ensuring scalability and reproducibility throughout the study.

\subsection{Antibody Responder Prediction Details}
\textbf{Data filtering and sample filtering.} We filtered studies based on biological relevance, retaining only influenza vaccine studies that included antibody measurements. Studies with extremely unbalanced responder distributions were excluded to ensure reliable model training. Additionally, we restricted the dataset to samples collected within the early immune response window (before day 28) and removed samples from control or placebo arms to focus on true vaccine-induced responses.

\textbf{Normalization and batch effect correction.} For RNA-seq samples, we applied a log$_2$(x + 1) transformation to stabilize variance. Genes with low signal—specifically those in the lowest 25\% of both mean expression and variance—were removed to reduce noise. Batch effects were addressed using three normalization strategies: Quantile Normalization (QN,\citealp{bolstad2003comparison}) mimicking the R \texttt{normalize.quantiles} method, ComBat~\citep{johnson2007adjusting} adjustment, and regression of batch effects using \texttt{scanpy}~\citep{wolf2018scanpy}.

\textbf{Padding.} For each subject, we created a time-ordered matrix combining gene expression with numeric timepoint values, and padded sequences to a uniform maximum length (maximum number of observations).

\textbf{PCA + GEE Baseline.}
To establish a population-averaged baseline for longitudinal prediction, we applied a generalized estimating equations (GEE) model to principal components (PCs) of the normalized gene-expression matrix. We computed the top 15 PCs—capturing the bulk (>\,80\%) of the variance—to reduce dimensionality and multicollinearity while retaining the dominant biological signal, and included the numeric timepoint as a covariate to account for temporal trends. Model performance was assessed with stratified 5-fold cross-validation repeated across six random seeds; test-set predictions were aggregated to the subject level via majority vote over visits. 

\textbf{RNN, LSTM, and GRU Models.} To capture temporal patterns in the longitudinal gene expression data, we implemented three deep learning sequence models: SimpleRNN~\citep{elman1990finding,rumelhart1986learning}, LSTM~\citep{hochreiter1997long}, and GRU~\citep{cho2014learning}. Each model was designed with a single recurrent layer of 64 hidden units followed by a dropout layer (rate = 0.4) to reduce overfitting and a final dense sigmoid output layer for binary classification. We used a masking layer to handle varying sequence lengths and padded all sequences to a unified maximum time length. The models were trained with the Adam optimizer~\citep{kingma2014adam} (learning rate = 1e-4), binary cross-entropy loss, and a batch size of 64 for 20 epochs. To ensure robust evaluation, stratified 5-fold cross-validation was repeated across six random seeds. These hyperparameters were chosen based on common practice in biomedical time-series modeling and to maintain consistency across architectures for fair comparison across batch correction methods (QN, ComBat, and Regression).

\textbf{Transformer Model.} 
To capture temporal patterns and feature interactions in longitudinal gene expression data, we implemented a custom Transformer architecture~\citep{vaswani2017attention}. 
At each visit, gene expression features were concatenated with the numeric timepoint, projected through a dense layer, and passed to a MultiHeadAttention mechanism (4 heads, key dimension 64) followed by layer normalization. 
The resulting sequence representations were aggregated via global average pooling and fed into dense layers with dropout for regularization. 
A final sigmoid output layer performed binary classification to predict immune response. 
The model was trained using the Adam optimizer (learning rate $1\times 10^{-4}$), binary cross-entropy loss, a batch size of 16, and up to 100 epochs. Stratified 5-fold cross-validation was repeated across six random seeds to ensure robust evaluation.

\textbf{t-SNE Details.} We run t-SNE using the \texttt{TSNE} class from the \texttt{sklearn.manifold} \citep{scikit-learn} module in Python. The input data matrix \texttt{X} (typically with shape: samples $\times$ features) is transformed into a 2-dimensional space using \texttt{fit\_transform(X)}. Key parameters include \texttt{n\_components=2} for 2D embedding, \texttt{perplexity=30} to balance local and global data structure, \texttt{max\_iter=1000} for optimization convergence, and \texttt{random\_state=42} for reproducibility. The resulting output is a 2D array where each row corresponds to the t-SNE embedding of the original sample, preserving the original row indexing.

\subsection{\textit{x}Cell Details}
We used xCell~\citep{aran2017xcell}, a gene signature-based method, to estimate the relative enrichment scores of 64 immune and stromal cell types from bulk transcriptomic data. xCell integrates gene expression signatures with spillover compensation to deconvolute cellular heterogeneity in bulk samples, making it especially suitable for inferring cell-type activity in immune-related studies.

In our analysis, normalized bulk gene expression matrices were used as input to the \texttt{xCell} R package (v1.1.0). The xCell scores were computed using the default pipeline, which includes transformation of gene expression values, enrichment score calculation using single-sample gene set enrichment analysis (ssGSEA, \citealp{barbie2009systematic}), and subsequent spillover compensation. The resulting cell-type enrichment scores range from 0 to 1 and reflect the relative abundance or activity of each cell type within the bulk sample.

To define a representative score for each immune cell type, we grouped the 64 xCell cell types into broader immune categories (e.g., B cells, T cells, NK cells, monocytes). For each group, we computed the mean xCell enrichment score across the constituent subtypes. Since the xCell enrichment scores for the 64 cell types are not mutually exclusive and often exhibit overlap in gene signatures, we did not use the sum of the enrichment scores to approximate overall immune cell proportions, as is commonly done in methods that produce compositional outputs. For example, the T cell score was defined as the average of the xCell scores for CD4+ naive T cells, CD4+ memory T cells, CD8+ T cells, regulatory T cells, and other annotated T cell subsets. This grouping strategy enabled a more interpretable summary of immune landscape variations across samples.

\subsection{Single-Cell Annotation Details}
\subsubsection*{SingleR-Based Annotation}

To annotate major immune cell types in the scRNA-seq datasets, we used the \texttt{SingleR} package \cite{aran2019reference}, which assigns cell types to single cells by comparing their expression profiles to reference transcriptomic datasets. We used the \texttt{HumanPrimaryCellAtlasData()} from the \texttt{celldex} package as the reference. The Seurat object was converted to a \texttt{SingleCellExperiment} format using \texttt{as.SingleCellExperiment()}, and annotation was performed with the following call:

{\small
\begin{verbatim}
predictions <- SingleR(test = sce, ref = ref, labels = ref$label.main)
\end{verbatim}
}

Predicted labels were stored in the \texttt{\$SingleR\_labels} field of the Seurat object. The process was applied independently to each dataset (e.g., GSE201534, GSE246937), and visualizations were generated using \texttt{DimPlot()}.

\subsection*{GPTCelltype Annotation}

We used the \texttt{GPTCelltype} \cite{hou2024gptcelltype} to assign high-level cell annotations based on multi-resolution clustering and GPT-4-based label prediction. The method leverages large language models (LLMs) to interpret differentially expressed genes and return biologically coherent labels. The annotation pipeline involves the following steps. For each cell cluster, we identified the top differentially expressed genes using the Wilcoxon rank-sum test. The top 10 marker genes per cluster were selected based on adjusted p-values and log fold changes. We use the model \texttt{gpt-4o} and a tissue-specific prompt: ``human peripheral blood mononuclear cells (PBMCs)''. The example function call is as below:

{\small
\begin{verbatim}
annotation_result <- GPTCelltype::gptcelltype(markers, model = "gpt-4o", 
tissuename = "human peripheral blood mononuclear cells (PBMCs)")
\end{verbatim}
}

\subsubsection*{Merging SingleR and GPTCelltype Annotations}

To ensure biological validity and avoid overannotation, we applied a harmonized approach. Low-confidence or rare SingleR annotations (defined as cell types with fewer than 100 cells) and biologically implausible labels were removed. Missing or ambiguous annotations were replaced by GPTCelltype labels. The example function call is as follows:

{\small
\begin{verbatim}
low_count_labels <- names(table(obj$SingleR_labels)[table(obj$SingleR_labels) < 100])
obj$cell_type <- obj$SingleR_labels
obj$cell_type[obj$SingleR_labels %in% low_count_labels] <- NA
obj$cell_type[is.na(obj$cell_type)] <-obj$celltype_res.0.1[is.na(obj$cell_type)]
\end{verbatim}
}
Final labels were stored in the \texttt{\$cell\_type} field and used for downstream proportion and visualization analyses (e.g., UMAPs and bar plots). A final harmonization step was conducted to unify terminology across datasets.

\subsubsection*{UMAP Visualization}

We utilized the \texttt{Seurat} R package \cite{hao2023dictionary} for dimensionality reduction and visualization. Uniform Manifold Approximation and Projection (UMAP) embeddings were computed using the \texttt{RunUMAP()} function, following principal component analysis (PCA) for initial dimensionality reduction. Annotated plots were generated with \texttt{DimPlot()}, enabling visualization of distinct cell clusters. To enhance the interpretability of these plots, we applied color palettes from the RColorBrewer package \cite{RColorBrewer}.

\bigskip

\bibliography{bibliography}

\end{document}